\documentclass[12pt]{iopart}
\usepackage{epsf,here}

\def \bea {\begin{eqnarray}}
\def \ea {\end{eqnarray}}
\def \vector#1{\stackrel{\hspace{-0.45em}\longrightarrow}{#1}}
\def \no {\nonumber}
\def \sub {\scriptscriptstyle}
\def\beq{\begin{equation}}
\def\eeq{\end{equation}}
\def\beqar{\begin{eqnarray}}
\def\eeqar{\end{eqnarray}}
\def\barr#1{\begin{array}{#1}}
\def\earr{\end{array}}
\def\bfi{\begin{figure}}
\def\efi{\end{figure}}
\def\btab{\begin{table}}
\def\etab{\end{table}}
\def\bce{\begin{center}}
\def\ece{\end{center}}

\def\text{\textstyle}

\begin{document}

\title[Anomalous Couplings]{Anomalous triple and quartic gauge boson
couplings} 

\author{P~J~Dervan$^1$, A~Signer$^2$, W~J~Stirling$^{2,3}$, 
A~Werthenbach$^2$} 

\address{$^1$ Department of Physics and Astronomy, University College,
London, England}

\address{$^2$ Department of Physics, University of Durham, Durham DH1
3LE, England}

\address{$^3$ Department of Mathematical Sciences, University of
Durham, Durham DH1 3LE, England}

\begin{abstract}
This article reviews some recent developments in the analysis of
anomalous triple and quartic vector boson couplings that have been
discussed at the UK Phenomenology Workshop on Collider Physics 1999 in
Durham.
\end{abstract}


\section{Introduction}

Triple and quartic gauge boson couplings arise in the Standard Model
due to the non-abelian nature of the theory and are, therefore, a
fundamental prediction. The study of these couplings is mainly
motivated by the hope that some new physics may result in a
modification of the couplings. If the new physics occurs at an energy
scale well above that being probed experimentally, it is possible to
integrate it out. The result is an effective theory with unknown
coefficients of the various operators in the Lagrangian. Any theory
beyond the Standard Model should be able to predict these
coefficients.  However, lacking a serious candidate for such a theory,
they are simply treated as anomalous couplings.

In most analyses of the past (see e.g. reference~\cite{EW98} for a review)
the Lagrangian was required to conserve $C$ and $P$ separately. This
was mainly motivated by the fact that this requirement leads to a
reduction of unknown parameters. Furthermore, the analyses
concentrated on triple gauge boson couplings, since the obtainable
constraints for these couplings are much stronger. It is not the aim
of this article to review these analyses, but rather to present two
new possibilities to look for anomalous couplings. The first is
concerned with the search for $CP$-violating triple gauge boson
couplings and the second with anomalous quartic couplings. Both are
done for LEP and they will be discussed in sections~\ref{anom:CPviol}
and \ref{anom:quarticLEP} respectively.

An extension of the study of anomalous quartic couplings to the
Tevatron is presented in section~\ref{anom:quarticTEV}. In the case of
a hadron collider, the search for anomalous couplings is complicated
by the fact that form factors have to be introduced. Indeed, since the
inclusion of anomalous couplings spoils the gauge cancellation in the
high energy limit, the effective theory will lead to violation of
unitarity for increasing partonic center of mass energy $\hat{s}$. In
order to prevent this, a suppressing factor is needed. The form of
this factor and the scale of new physics associated with it is to a
large extent arbitrary. This introduces an unwanted dependence of
limits on anomalous couplings on the precise form of the form
factor. In section~\ref{anom:Had} we investigate to which extent this
arbitrariness in measurements of anomalous couplings at hadron
colliders can be avoided.

\section{$CP$-violating gauge boson couplings at LEP \label{anom:CPviol}}

The study of the helicity of the intermediate state $W$ bosons gives
direct access to a model independent test of the Standard Model. The
values of both $CP$ conserving and $CP$ violating anomalous Trilinear
Gauge boson couplings can be directly measured by comparing the W
helicity properties with those predicited in the Standard Model.

According to the most general Lorentz invariant
Lagrangian~\cite{dieter,bilenky}, there are 14 independent couplings
describing the WWV vertex $( { \rm V} = {\rm Z},\gamma)$.  Within the
$SU(2)_{L} \times U(1)_{Y}$ theory, $CP$ violation is only present in
the ${ \rm e}^{+}{\rm e}^{-} \rightarrow {\rm W}^{+}{\rm W}^{-}$
process via the Kobayashi-Maskawa phase~\cite{koby}, which affects it
at the two loop order only. $CP$ violating terms for the trilinear
$\gamma$WW and ${\rm Z}^{0}$WW interactions are, however, easily
included in the $SU(2)_{L} \times U(1)_{Y}$
Lagrangian~\cite{dieter,gounaris}. There are then 4 couplings which
violate $P$ and $CP$ invariance, $\tilde{\kappa}_{V}$ and
$\tilde{\lambda}_{V}$, and 2 which violate $C$ and $CP$ invariance
$g^{V}_{4}$. Within the Standard Model all these couplings are zero 
but a linear realization of the basic $SU(2)_{L}\times U(1)_{Y}$ 
symmetry gives the following relations~\cite{papa} between the $CP$ 
violating couplings:
\beq
\tilde{\kappa}_{Z}=-\tan^{2}\theta_{w}\tilde{\kappa}_{\gamma}; \quad
\tilde{\lambda}_{Z}=\tilde{\lambda}_{\gamma}; \quad
g^{Z}_{4} = g^{\gamma}_{4}
\eeq

One method of measuring the $CP$-violating couplings is the Spin Density
Matrix (SDM) analysis. The two-particle joint SDM~\cite{bilenky}
completely describes the helicity of the $W$ bosons produced in the
triple gauge boson interaction.  The matrix elements are observables
directly related to the polarisation of the $W$ bosons and so their
measurement will give direct access to the underlying physics of the
WW production process and allow a model independent test of the TGCs.
This method of analysis is extremely desirable for investigating the
$CP$-violating couplings, because a number of the SDM elements'
coefficients are particularly sensitive to the $CP$-violating couplings
while being unaffected by changes in the $CP$-conserving couplings.

The matrix elements are normalised products of the helicity amplitudes
of the $W^{+}$ and $W^{-}$. The matrix is hermitian and therefore has
80 independent elements if the off-diagonal elements are complex.
This results in 80 independent coefficients to be experimentally
measured.  The diagonal elements are purely real and are equivalent to
the probability of producing a final WW state with helicity $\tau_{-}
\tau_{+}$ (where $\tau_{-}$ and $\tau_{+}$ are the helicity states of
the $W^{+}$ and $W^{-}$).  The off-diagonal elements are the cross
terms from the interference of all the possible final states. The
number of independent elements can be further reduced by only
considering the decay of one $W$ and summing over all possible
helicity states of the other.  This single $W$ SDM has only nine
elements.

The single $W$ SDM matrix is hermitian, the off-diagonal components of
which are once again complex, leading to the nine independent SDM
coefficients. The diagonal elements are real and can be interpreted as
the probability of producing a $W$ boson of the respective helicity,
$\tau$. Therefore, they are normalised to unity.
The imaginary SDM coefficients are extremely sensitive to $CP$ violation
at the three gauge boson vertex but completely insensitive to $CP$
conserving anomalous couplings.  However, in a theory with no $CP$
violation at the vertex, any deviation from zero in the imaginary SDM
coefficients could only be due to loop effects.

The unnormalised single $W$ SDM elements can be extracted from the data
of the decay product angles by integrating with suitable spin
projection operators~\cite{operate} that reflect the standard V-A
coupling of fermions to the $W$ boson in the $W$ decay.  

The theoretical predictions for the single $W$ SDM elements as a
function of anomalous couplings can be derived from the analytical
expressions of the helicity amplitudes~\cite{operate}.  The single $W$
matrix elements can be extracted using the three-fold differential
cross section~\cite{gounaris,operate}.  This extraction method uses
the data event by event, so each event is analysed individually and
then the sum of all events in the bin is taken.

Certain projection operators are symmetric under the transformation
$\cos\theta^{*} \rightarrow -\cos\theta^{*}$, $\phi^{*} \rightarrow
\phi^{*} + \pi$ and so a number of the SDM elements (or combinations
thereof), can be extracted from the folded angular distribution of the
hadronically decaying $W$ in the semileptonic event, where
differentiation between the particle and anti-particle decay product
is extremely difficult.  

  \vspace*{-1cm}
  \begin{figure}[htbp]
  \centerline{\epsfysize=10.0 cm\epsffile{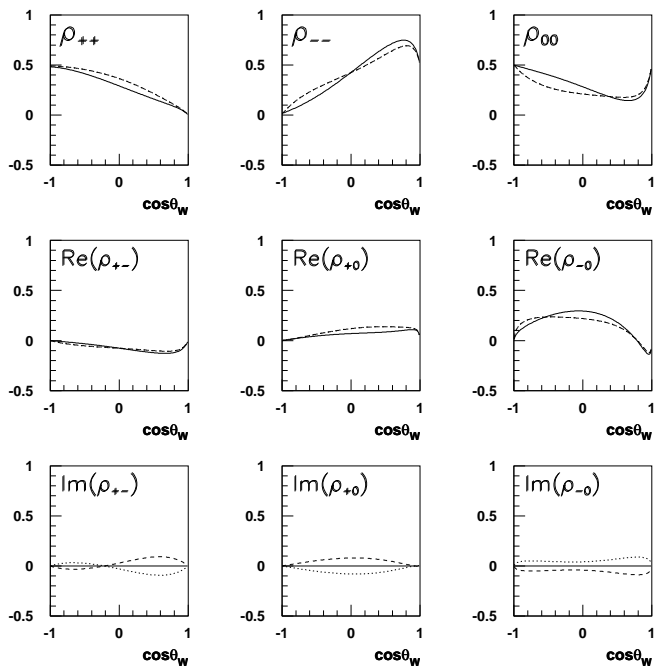}}  
  \vspace*{-0.5cm}
  \caption{\label{cpelements}{The SDM elements for the Standard Model
  (solid), $\tilde{\lambda}_Z = +1$ (dotted) and $\tilde{\lambda}_Z=
  -1$ (dashed).} }
  \end{figure}

Figure~\ref{cpelements} shows the SDM elements for the Standard
Model (solid), $\tilde{\lambda}_Z = +1$ (dotted) and $\tilde{\lambda}_Z
= -1$ (dashed).
This also means using suitable combinations of projection operators,
certain combinations of the joint particle SDM elements,  can be 
extracted from the 5 fold differential cross section. 

\begin{figure}[htbp]
  \centerline{\epsfysize=11cm\epsffile{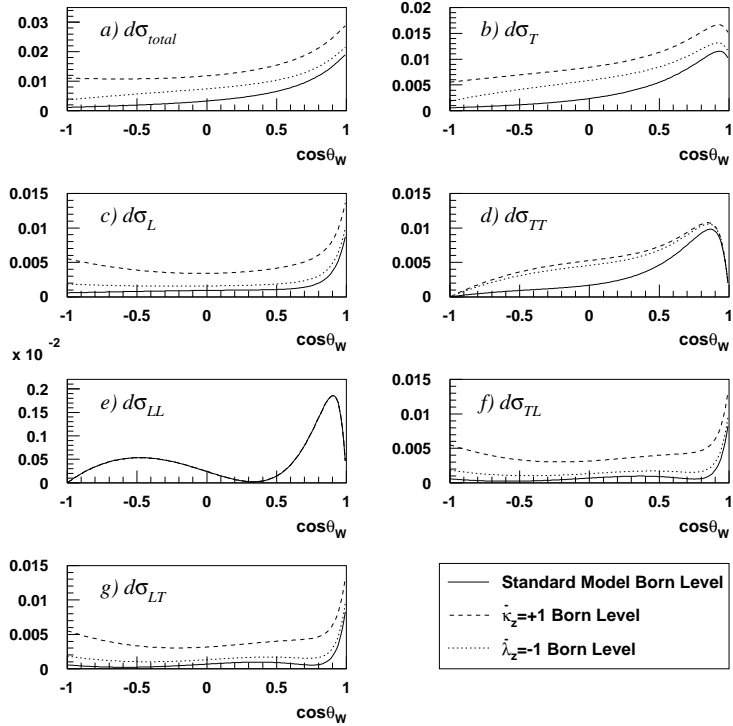}}
  \vspace*{-0.5cm}
  \caption{  \label{helicity}{
   Polarised differential cross sections for the Standard
   Model (solid), $\tilde{\kappa}_Z = +1$ (dotted) and
   $\tilde{\kappa}_Z = -1$ (dashed).} }
  \end{figure}

The SDM elements are directly related to the polarisation of the W
bosons, so they can be used to extract the polarised differential
cross sections from the data. Figure~\ref{helicity} shows the above 
polarised differential cross section for the Standard Model (solid), 
$\tilde{\kappa}_Z = +1$ (dotted) and $\tilde{\kappa}_Z = -1$ (dashed).
  \vspace*{-1cm}

A study of the $CP$-violating couplings has been performed at OPAL using W 
pair events which decay semileptonically, from the data recorded in 
1998, at a centre-of-mass energy of 189 GeV with an integrated
luminosity of 183 pb$^{-1}$.

\section{Anomalous quartic couplings at LEP \label{anom:quarticLEP}}

The lowest dimension operators which lead to genuine quartic couplings 
where at least one photon is involved are of dimension 6 \cite{belanger}.
The two most commonly studied are \cite{us}
\bea
\label{L0}
{\cal L}_0 &=& - \frac{e^2}{16 \Lambda^2}\, a_0\, F^{\mu \nu} \, F_{\mu
\nu} \vector{W^{\alpha}} \cdot \vector{W_{\alpha}} \no \\
&=&  - \frac{e^2}{16 \Lambda^2}\, a_0\, \big[ - 2 (p_1 \cdot p_2 ) 
( A \cdot A) + 2 (p_1 \cdot A)(p_2 \cdot A)\big] \no \\
&& \hspace{1.5cm} {\sub \times} \big[ 2 ( W^+ \cdot W^-) +  (Z \cdot Z) /
\cos ^2 \theta_w \big]  \quad ,
\ea
\bea
\label{Lc}
{\cal L}_c &=& - \frac{e^2}{16 \Lambda^2}\, a_c\, F^{\mu \alpha} \, F_{\mu
\beta} \vector{W^{\beta}} \cdot \vector{W_{\alpha}} \no \\
&=& - \frac{e^2}{16 \Lambda^2}\, a_c\, \big[- (p_1 \cdot p_2)\, A^{\alpha}
A_{\beta} +(p_1 \cdot A)\, A^{\alpha} p_{2 \beta} \big. \no \\
&& \hspace{1.8cm}\big. \quad \quad + (p_2 \cdot A)\,
p_1^{\alpha} A_{\beta} -(A \cdot A)\, p_1^{\alpha} p_{2 \beta} \big] \no
\\
&& \hspace*{1.5cm} {\sub \times} \big[ W_{\alpha}^- W^{+ \beta} +
W_{\alpha}^+ W^{-
\beta} + Z_{\alpha} Z^{\beta} / {\cos ^2 \theta_w} \big] \ . 
\ea
both giving anomalous contributions 
to the $VV\gamma\gamma$ vertex, with $VV$ either being $W^+W^-$ or
$Z^0Z^0$, where $p_1$ and $p_2$ are the photon momenta and
\bea
\label{wb}
\vector{W _{\mu}} = \left( \begin{array}{c}  \frac{1}{\sqrt{2}} (
W_{\mu}^+ 
+ W_{\mu}^-) \\  \frac{i}{\sqrt{2}} ( W_{\mu}^+ - W_{\mu}^-) 
\\  \frac{Z_{\mu}}{\cos \theta_w}  \end{array} \right) \:.
\ea

Anomalous $VZ\gamma\gamma$ vertices can in principle also be
considered.  Since the sensitivity to those is much smaller we
restrict ourselves in this analysis to the two anomalous parameters
$a_0$ and $a_c$. For a complete set of operators of this type see
Ref.~\cite{eboli}.

The anomalous scale parameter $\Lambda$ that appears in the above
anomalous contributions has to be fixed.  In practice, $\Lambda$ can
only be meaningfully specified in the context of a specific model for
the new physics giving rise to the quartic couplings.  However, in
order to make our analysis independent of any such model, we choose to
fix $\Lambda$ at a reference value of $M_W$, following the conventions
adopted in the literature. Any other choice of $\Lambda$
(e.g. $\Lambda = 1$~TeV) results in a trivial rescaling of the
anomalous parameters $a_0$ and $a_c$.

It follows from the Lagrangian that any anomalous contribution is {\it
linear} in the photon energy $E_{\gamma}$. This means that it is the
hard tail of the photon energy distribution that is most affected by
the anomalous contributions, but unfortunately the cross sections here
are very small.  In the following numerical studies we will impose a
lower energy photon cut of $E_{\gamma}^{\rm min} = 20$~GeV. Similarly,
there is also no anomalous contribution to the initial-state photon
radiation, and so the effects are largest for centrally-produced
photons. We therefore impose an additional cut of $\vert
\eta_\gamma\vert < 2$\footnote{Obviously in practice these cuts will
be tuned to the detector capabilities.}.  We do not include any
branching ratios or acceptance cuts on the decay products of the
produced $W^\pm$ and $Z^0$ bosons, since we assume that at $e^+e^-$
colliders the efficiency for detecting these is high. 

\begin{figure}[h]
\centerline{\epsfysize=8cm\epsffile{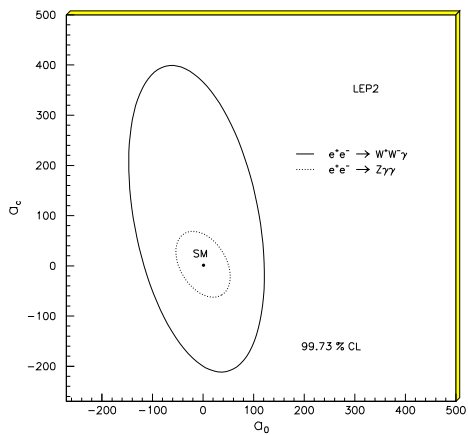}}
\vspace{-1.5cm}
\caption{\label{com200}{Contour plots for $+3 \, \sigma$ deviations
from the SM $e^+e^-\to W^+W^-\gamma$ and $e^+e^-\to Z^0\gamma\gamma$
total cross sections at $\sqrt{s} = 200$~GeV with $ \int {\cal L} =
130$~pb$^{-1}$. }}
\end{figure}

Figure~\ref{com200} shows the contour in the $(a_0,a_c)$ plane that
corresponds to a $+3 \sigma$ deviation of the $WW\gamma$ and
$Z^0\gamma\gamma$ SM cross sections at $\sqrt{s} = 200$~GeV with $\int
{\cal L} = 150$~pb$^{-1}$.

The key features in determining the sensitivity for a given process,
apart from the fundamental process dynamics, are the available photon
energy $E_{\gamma}$, the ratio of anomalous diagrams to SM
`background' diagrams, and the polarisation state of the weak bosons
\cite{belanger}. A high-energy linear collider ($\sqrt{s} \sim 500 -
1000$~GeV), would allow more phase space for photon emission, and
would give significantly tighter bounds on the coupling, see
Ref.~\cite{us}.  At LEP2 energies $Z^0\gamma\gamma$ benefits
kinematically from producing only one massive boson, which leaves more
energy for the photons as well as having fewer `background'
diagrams. On the other hand $W^+W^-\gamma$ production at this
collision energy suffers from the lack of phase space available for
energetic photon emission, although this is partially compensated by
the production of longitudinal bosons, which gives rise to higher
sensitivity to the anomalous couplings.

Finally, it is important to emphasise that in our study we have only
considered `genuine' quartic couplings from new six-dimensional
operators. We have assumed that all other anomalous couplings are
zero, including the trilinear ones.  Since the number of possible
couplings and correlations is so large, it is in practice very
difficult to do a combined analysis of\protect{ {\it all}} couplings
simultaneously.  In fact, it is not too difficult to think of new
physics scenarios in which effects are only manifest in the quartic
interactions. One example would be a very heavy excited $W$ resonance
produced and decaying as in $W^+ \gamma \to W^* \to W^+ \gamma$.

\vspace*{-1.cm}
%

\section{Anomalous quartic couplings at the Tevatron
\label{anom:quarticTEV}} 

\begin{figure}[H]
\centerline{\epsfysize=8cm\epsffile{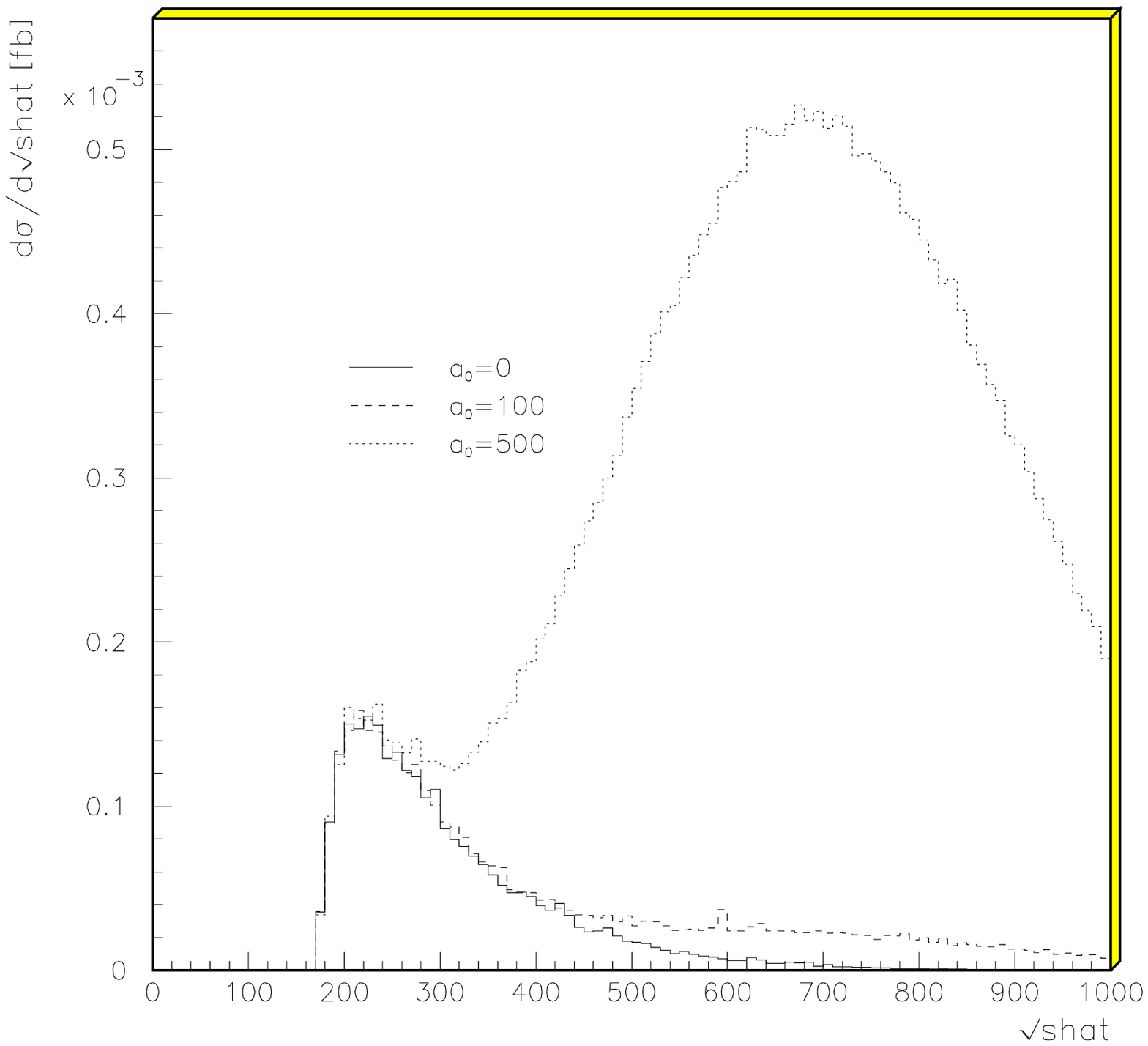} \hspace*{1cm}
            \epsfysize=8cm\epsffile{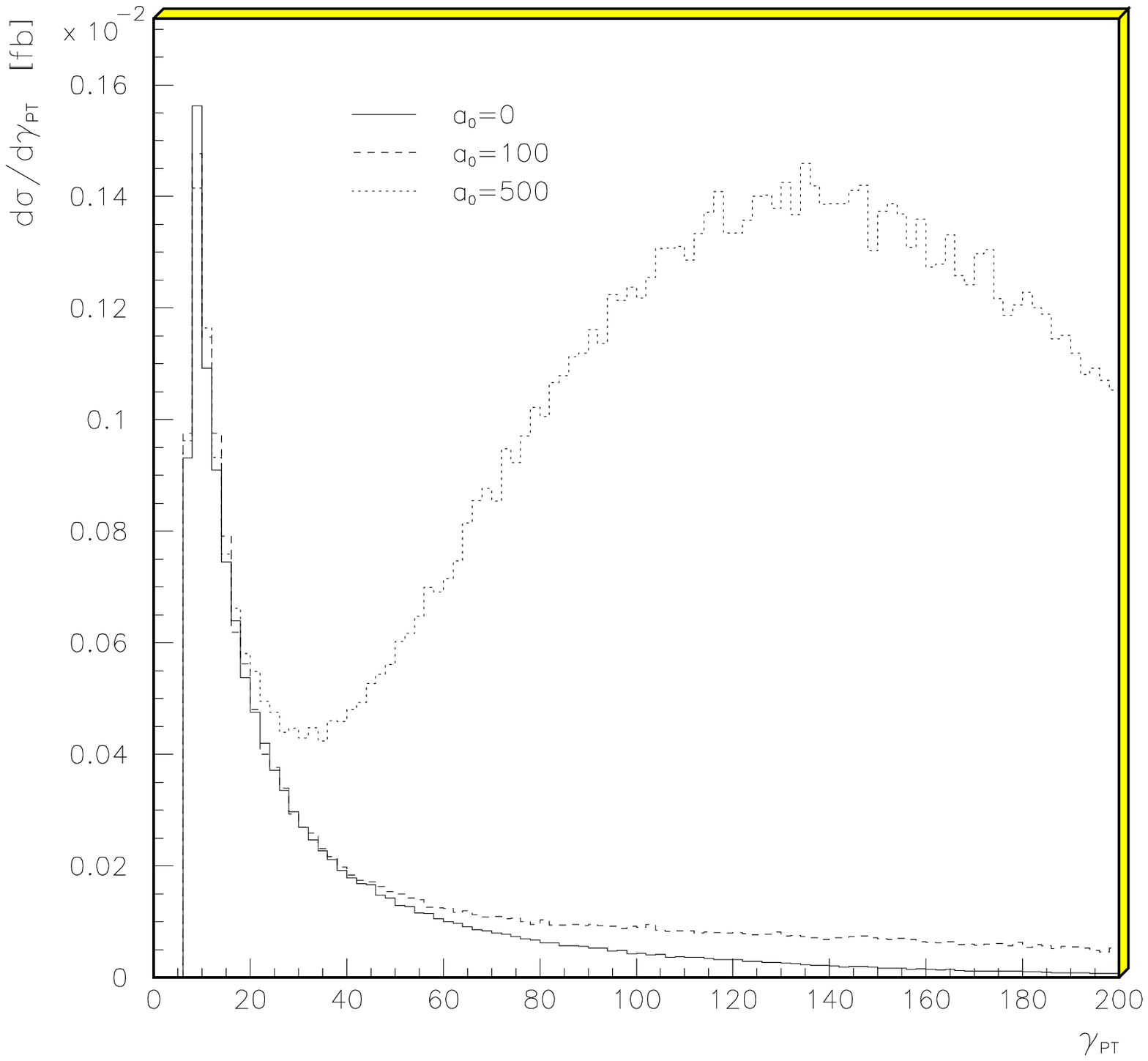}}
\vspace{-1.5cm}
\caption{\label{shat}{$\sqrt{\hat{s}}$ spectrum (left) and transverse
photon energy spectrum (right) for the process $p\bar{p} \to
W^+W^-\gamma$ in the presence of the anomalous coupling $a_0$.}}
\end{figure}

Motivated by a request from experimentalists at the Workshop, we
investigated the sensitivity of the processes $ p\bar{p} \to
W^+W^-\gamma$ and $Z\gamma\gamma$ to the above anomalous quartic
couplings, $a_0$ and $a_c$.  We consider a Tevatron scenario of
$\sqrt{s}=2$~TeV with an integrated luminosity $\int {\cal
L}=2$~fb$^{-1}$ and impose a transverse momentum cut $p_{\perp \gamma}
> 10$~GeV and a rapidity cut of $\vert \eta_\gamma\vert <2.5$ on the
final-state photon(s).  It can be seen from Figure~\ref{shat} that the
mean partonic centre-of-mass energy is $\sim 250$~GeV and hence it is
possible to perform the analysis without the need to introduce a form
factor. For ease of comparison with the LEP results, we again choose
the anomalous scaling parameter $\Lambda=M_W$. 

For purposes of illustration we only consider here the sensitivity of
the cross sections to one of the anomalous parameters, $a_0$, since
this one has the highest sensitivity.  Thus Figure~\ref{shat} shows
the partonic centre-of-mass spectrum corresponding to $a_0 = 0,\ 100,\
500$, with $a_c=0$.  Again for the purpose of illustration we have
chosen here to display the results for the process $ W^+W^-\gamma$
only. Similar results are found for $Z\gamma\gamma$ production.

\vspace{-1.cm}

In Figure~\ref{shat} we also show the impact of the anomalous
parameter $a_0$ on the transverse momentum of the photon. As
anticipated above, it is the hard tail of the photon spectrum that is
particularly sensitive to the anomalous contributions and this
observable therefore offers a means to search directly for such
anomalous contributions.

\begin{figure}[H]
\centerline{\epsfysize=8cm\epsffile{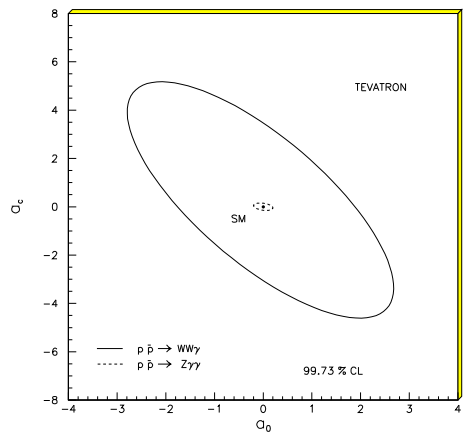}}
\vspace{-1.5cm}
\caption{\label{com}{Contour plots for $+3 \, \sigma$ deviations
from the SM $p\bar{p}\to W^+W^-\gamma$ and
$p\bar{p}\to Z^0\gamma\gamma$
total cross sections at $\sqrt{s} = 2$~TeV
with $ \int {\cal L} = 2$~fb$^{-1}$.}}
\end{figure}

Finally, we have studied the impact on the {\it total} cross sections
of the processes $ p\bar{p} \to W^+W^-\gamma$ and
$Z\gamma\gamma$. Figure~\ref{com} shows the contour in the $(a_0,a_c)$
plane corresponding to the $+3\sigma$ deviation of the SM cross
section. Just as in the LEP2 study, $Z\gamma\gamma$ production
promises a better discovery potential, again due to the higher photon
energy available.

\vspace{-1.cm}

Comparing Figures~\ref{com200} and \ref{com}, we conclude that the
Tevatron should be able to set much tighter limits on the anomalous
couplings $a_0$ and $a_c$ than LEP. The basic difference is the
significantly higher subprocess scattering energies available at the
hadron collider.

\section{Measurements of anomalous couplings at hadron colliders
\label{anom:Had} } 

At $e^+e^-$ colliders anomalous couplings are directly investigated at
a fixed centre of mass energy $\sqrt{s}$. This results in bounds of
the anomalous couplings $\alpha_{\rm ac}$ as a function of $s$. At
hadron colliders, the center of mass energy of the colliding partons
$\sqrt{\hat{s}}$ is not fixed and there will be events where
$\sqrt{\hat{s}}$ is very large. In order to avoid problems with
violation of unitarity, form factors $f$ are introduced, i.e. the
anomalous couplings $\alpha_{\rm ac}$ are replaced by $\alpha_{\rm ac}
f$. The precise form of $f$ as well as the associated scale for new
physics $\Lambda$ are to a big extent arbitrary. A common choice is
\begin{equation}
f = \left(1+\frac{\hat{s}}{\Lambda^2}\right)^{-n}
\end{equation}
where $n$ is chosen big enough to ensure unitarity for
$\hat{s}\to\infty$.  This procedure has the unpleasant consequence
that all bounds on the anomalous couplings depend on $\Lambda$ and the
precise form of $f$. 

In order to improve the situation it is desirable to get bounds
directly on $\alpha_{\rm ac}(\sqrt{\hat{s}})$ also at hadron
colliders. In some cases, this is straightforward to do. As an example
we mention $Z\gamma$ production which probes the anomalous couplings
$h_3^\gamma, h_4^\gamma, h_3^Z$ and $h_4^Z$. In this process $\hat{s}$
can be fully reconstructed. This allows to investigate the anomalous
couplings in different regions of $\hat{s}$ and get separate bounds in
each region. At the LHC the statistics should be good enough to allow
such an analysis.

\begin{figure}[h]
\centerline{\epsfysize=14cm\epsffile{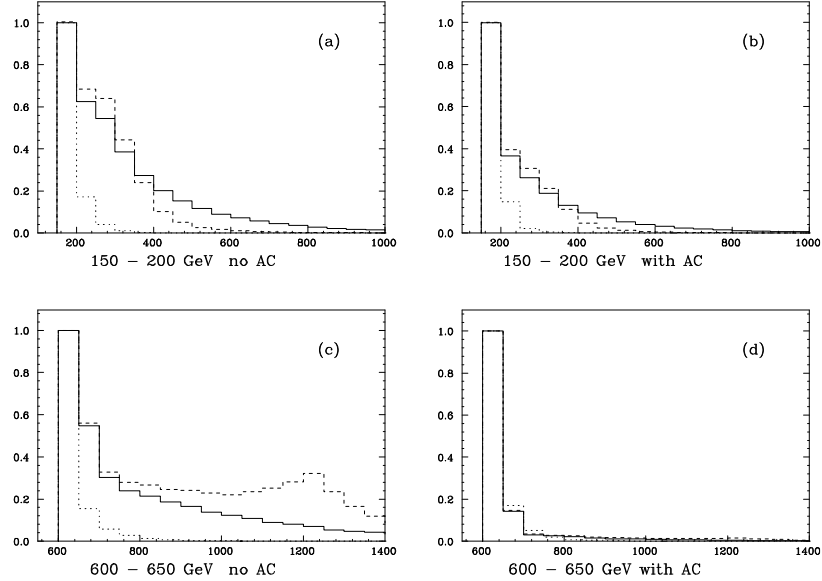}}
\vspace{-5.5cm}
\caption{\label{corr}{Correlation of $M_T$ (solid line), $M_C$ (dashed
line) and $\hat{s}_{\rm min}$ (dotted line) with $\hat{s}$ for two
bins in the observable for the LHC. Plots (a) and (c) are Standard
Model results whereas plots (b) and (d) include large anomalous
couplings. }}
\end{figure}

The situation is more difficult in processes, where $\hat{s}$ can not
be fully reconstructed, such as $pp\to W\gamma \to \ell\nu\gamma
X$. In these cases an observable quantity has to be found which has a
very strong correlation to $\hat{s}$. Then the analysis could be done
using this quantity instead of $\hat{s}$ without introducing a large
error.  There are several possibilities, such as the transverse mass
$M_T$ or the cluster mass $M_C$. They are defined as follows:
\begin{eqnarray}
M_T^2 &\equiv& \biggl( \sqrt{p_T^2(\ell\gamma)+m^2(\ell\gamma)} +
|p_T(\nu)| \biggl)^2 - p_T^2(\ell\gamma\nu)  \\
M_C^2 &\equiv& \biggl(
\sqrt{p_T^2 (\ell\gamma)+m^2 (\ell\gamma)} + |p_T(\nu)| \biggl)^2 
\end{eqnarray}
Note that at leading order $M_T = M_C$. Another possibility is to take
$\hat{s}_{\rm min}$ \cite{dFKS99} which is defined as follows:
assuming the $W$ to be on-shell and identifying the missing transverse
momentum with $p_T(\nu)$ it is possible to reconstruct the full
kinematics with a twofold ambiguity. This result in two possible
values of $\hat{s}$ and $\hat{s}_{\rm min}$ is by definition the
smaller of the two. In Figure~\ref{corr} the distribution of the true
$\hat{s}$ is shown for two particular bins of the observed
quantity. The curves have been obtained for $pp$ collision at
$\sqrt{s}$=14~TeV with some appropriate cuts on the rapidity and
transverse momentum of the leptons. For plots (a) and (b) we have
$150~{\rm GeV} < {\cal Q} < 200~{\rm GeV}$ whereas for plots (c) and
(d) we have $600~{\rm GeV} < {\cal Q} < 650~{\rm GeV}$ where the
observed quantity ${\cal Q} \in \{M_T, M_C,\hat{s}_{\rm min}\}$. To
make the comparison of the various correlations easier, the histogram
has been normalized to one in the first bin. The by far strongest
correlations are obtained for $\hat{s}_{\rm min}$. Even if we include
unrealistically large anomalous couplings the correlation is
preserved, if not enhanced. This can be seen in Figures~\ref{corr} (b)
and (d), where we show the correlations with $\Delta\kappa^\gamma =
0.8, \lambda = 0.2$ and the ususal dipole form factor with a scale
$\Lambda=1$~TeV. These results have been obtained with a Monte Carlo
program including next-to-leading order QCD corrections
\cite{dFKS99}. The large correlation between $\hat{s}$ and
$\hat{s}_{\rm min}$ should allow for a similar analysis as in the case
of $Z\gamma$ production by simply replacing $\hat{s}$ by $\hat{s}_{\rm
min}$.
\vspace{0.5cm}




\section*{References}
\begin{thebibliography}{99}
%
%
\bibitem{EW98} Ellison~J, Wudka~J 1998 {\it preprint} hep-ph/9804322.
%
%
\bibitem{dieter} Hagiwara~K et al. 1987 {\it Nucl.~Phys.} {\bf
B282} 253. 
\bibitem{bilenky} Bilenky~M, Kneur~J~L, Renard~F~M and
Schildknecht~D 1991 {\it Nucl. Phys.} {\bf 263} 291. 
\bibitem{koby} Kobayashi~M and Maskawa~M 1973 {\it Progr. Theor. Phys.} {\bf
49}, 652. 
\bibitem{gounaris} Gounaris~G, Schildknecht~D and Renard~F~M 1991
{\it Phys. Lett.} {\bf B263}, 291. 
\bibitem{papa} Gounaris~G, Papadopoulos~C~G 1998 {\it Eur.Phys.J.}
{\bf C2} 365.  
\bibitem{operate} Gounaris~G, Layssac~J, Moultaka~G and
Renard~F~M 1993 {\it Int.~J.~Mod.~Phys.} {\bf A8} 3285.
%
%
\bibitem{belanger} B\'elanger~G, Boudjema~F 1992 {\it
Phys.~Lett.} {\bf 288} 201.
\bibitem{eboli} \'Eboli~O~J~P, Gonzal\'ez-Garcia~M~C, Novaes~S~F  1994
{\it Nucl. Phys.} {\bf B411} 381; \\ 
\hspace*{-0.48cm}
B\'elanger~G et al. 1999 {\it preprint} hep-ph/9908254.
\bibitem{us} Stirling~W~J, Werthenbach~A   1999 {\it preprint}
hep-ph/9903315, to appear in {\it Euro.~Phys.~J.}
%
%
\bibitem{dFKS99} de~Florian~D and Signer~A 2000 {\it
preprint} hep-ph/0002138.  
%
\end{thebibliography}
\end{document}